\documentclass[aps,showpacs,two column]{revtex4}

\usepackage{graphicx}
\usepackage[squaren]{SIunits}

\usepackage{amssymb}
\usepackage{color}
\begin{document}
\title{Transport properties and superconductivity in K$_{1-x}$Ln$_{x}$Fe$_2$As$_2$ (Ln=Sm, Nd
and La) system}

\author{R. H. Liu, G. Wu, H. Chen, Y. L. Xie, Y. J. Yan, T. Wu, X. F. Wang,
J. J. Ying, Q. J. Li, B. C. Shi and X. H. Chen}
\altaffiliation{Corresponding author} \email{chenxh@ustc.edu.cn}
\affiliation{Hefei  National Laboratory for Physical Science at
Microscale and Department of Physics, University of Science and
Technology of China, Hefei, Anhui 230026, People's Republic of
China}

\begin{abstract}
We prepared the samples K$_{1-x}$Ln$_{x}$Fe$_2$As$_2$ (Ln=Sm, Nd and
La) with ThCr$_2$Si$_2$-type structure. These samples were
characterized by X-ray diffraction, resistivity, susceptibility and
thermoelectric power (TEP). Substitution of Ln (Ln=La, Nd and Sm)
for K in K$_{1-x}$Ln$_{x}$Fe$_2$As$_2$ system raises the
superconducting transition temperature to 34-36 K. The TEP
measurements indicate that the TEP of K$_{1-x}$Ln$_{x}$Fe$_2$As$_2$
is positive, being similar to the case of the
Ba$_{1-x}$K$_{x}$Fe$_2$As$_2$ system with p-type carrier. In the
K$_{1-x}$Ln$_{x}$Fe$_2$As$_2$ system, the superconducting
$KFe_2As_2$ with $T_c\sim 3$ K is the parent compound, and no
structural and spin-density wave instabilities exist in this system.
\end{abstract}
\vskip 15 pt  \maketitle

\section{Introduction}

The discovery of superconductivity at 26 K in LaO$_{1-x}$F$_{x}$FeAs
has generated tremendous interest in the field of high-T$_{C}$
superconductivity\cite{Kamihara}. LaFeAsO with the tetragonal
ZrCuSiAs-type structure becomes superconducting by doping to induce
carrier to the FeAs layers\cite{Kamihara}. Shortly after that
discovery, the T$_{C}$ was raised to 41-55 K by replacing La by
other rare-earth Ce\cite{gfchen1}, Sm\cite{xhchen,
 ren1,rhliu}, Pr\cite{ren2}, Nd\cite{ren3}, Gd\cite{cwang}, etc. The
crystal structure of LaOFeAs contains alternating layers of
edge-sharing Fe$_{2}$As$_{2}$ tetrahedral layers and La$_{2}$O$_{2}$
tetrahedral layers along c-axis. In addition to the
superconductivity\cite{Kamihara}, the system displays a closely
related electron-pair instability on the Fermi surface: the
spin-density-wave(SDW)\cite{ccruz}. The competition between
superconductivity and SDW instability was identified in the doped
LnOFeAs(1111) system\cite{rhliu,jlluo}.

 Recently, the ternary iron arsenide BaFe$_{2}$As$_{2}$ shows
 superconductivity at 38K by hole doping with partial substitution
 of potassium for barium\cite{rotter, gwu1}. There exists a single FeAs layer in
 an unit cell in LnOFeAs system, while there are two FeAs layers
 in an unit cell in BaFe$_{2}$As$_{2}$ with ThCr$_{2}$Si$_{2}$-type
 structure. The superconductivity is also observed at 37 K in
 isostructural K- and Cs-doped SrFe$_{2}$As$_{2}$\cite{cwchu, gfchen2}, at 20 K in
 Na-doped CaFe$_{2}$As$_{2}$\cite{gwu2} and at 32 K in K-doped EuFe$_{2}$As$_{2}$\cite{hsjeevan}.
 All these parent compounds share the common features of a structure
 transition and SDW ordering, which were characterized by the anomaly in
 resistivity, specific heat and magnetic susceptibility. Moreover,
 the structure transition and SDW ordering were suppressed, and the
 superconductivity was induced in the parent samples RFe$_{2}$As$_{2}$(R=Ca,Ba, Sr) under
 high pressure\cite{milton, alireza}. The superconductivity with T$_C\sim$ 3 K was
 found in isostructural KFe$_{2}$As$_{2}$, and no anomaly appears in
 resistivity\cite{cwchu, hchen}. It suggests that no structural and SDW transitions happen in KFe$_{2}$As$_{2}$.
The angle-resolved photoemission spectroscopy(ARPES) measurements on
the samples KFe$_{2}$As$_{2}$ and
Ba$_{0.6}$K$_{0.4}$Fe$_{2}$As$_{2}$ clarified key features of the
band structure and Fermi surface (FS) topology. Unlike
Ba$_{0.6}$K$_{0.4}$Fe$_{2}$As$_{2}$, the nesting condition via the
antiferromagnetism (AF) wave vector is not satisfied in
KFe$_{2}$As$_{2}$\cite{tsato}. If one takes KFe$_2$As$_2$ as a
parent compound, it is possible that substitution of the rare-earth
element for potassium induces more carrier into system and
consequently enhance superconducting transition temperature T$_C$.
Here we report the synthesis of the samples
K$_{1-x}$Ln$_{x}$Fe$_2$As$_2$ (Ln=Sm, Nd and La), and systematically
study their transport properties (resistivity and thermoelectric
power). It is found that substitution of Ln (La, Nd, Sm) for
potassium leads to an increase of T$_C$ up to 36 K. The behavior of
resistivity and TEP of K$_{1-x}$Ln$_{x}$Fe$_2$As$_2$ are very
similar to that observed in Ba$_{1-x}$K$_{x}$Fe$_2$As$_2$, and the
TEP is positive, indicating p-type carrier.

\begin{figure}[h]
\includegraphics[width=0.50\textwidth]{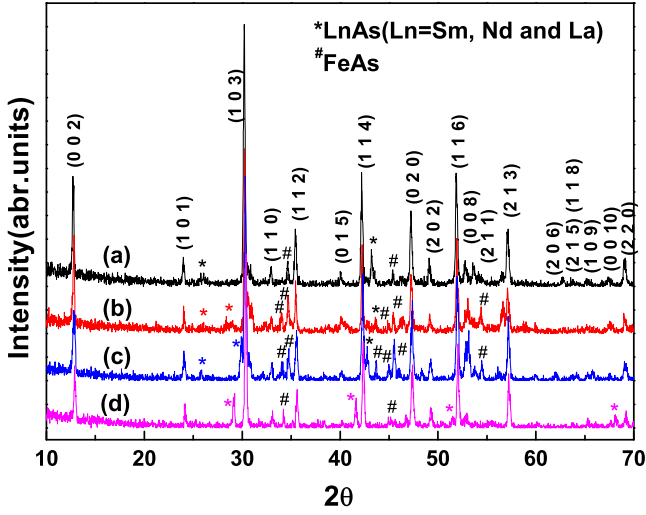}
\caption{X-ray powder diffraction patterns at room temperature for
the samples, (a): K$_{0.7}$Sm$_{0.3}$Fe$_2$As$_2$, (b):
K$_{0.6}$Sm$_{0.4}$Fe$_2$As$_2$, (c):
K$_{0.7}$Nd$_{0.3}$Fe$_2$As$_2$, (d):
K$_{0.7}$La$_{0.3}$Fe$_2$As$_2$.}
\end{figure}

Polycrystalline samples of K$_{1-x}$Ln$_{x}$Fe$_2$As$_2$ (Ln=Sm, Nd
and La) were synthesized by solid state reaction method using KAs,
Fe$_2$As and LnAs as starting materials. KAs was pre-synthesized by
heating the mixture of K lumps and As powder at 200$\celsius$ for 24
hours. LnAs was prepared by the reaction of Ln (La, Nd, Sm) powder
and As powder at 600$\celsius$ for 24 hours. The raw materials were
accurately weighed according to the stoichiometric ratio of
K$_{1-x}$Ln$_{x}$Fe$_2$As$_2$, then thoroughly grounded and pressed
into pellets. About 20-30\% excessive K was added because K was easy
to volatilize and reacted with quartz tube at high temperature. The
pellets were wrapped with Ta foil and sealed in quartz tube under
about 1/4 atm Ar gas atmosphere. The sealed tubes were sintered at
740 $\celsius$ for 24 hours. In order to improve its homogeneity and
stiffness, the resulting product was reground before applying a
second heat treatment at 720$\celsius$ for 20 hours. The sample
preparation process except for annealing was carried out in glove
box in which high pure argon atmosphere is filled.

\begin{figure}[t]
\includegraphics[width=0.50\textwidth]{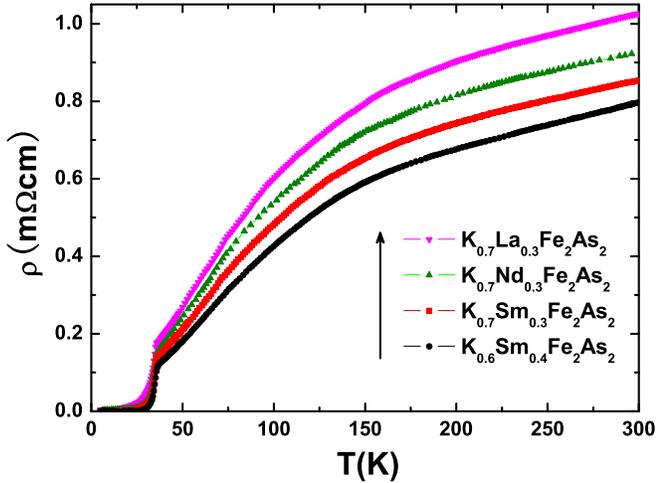}
\caption{Temperature dependence of resistivity for the samples, (a):
K$_{0.7}$Sm$_{0.3}$Fe$_2$As$_2$(squares); (b):
K$_{0.6}$Sm$_{0.4}$Fe$_2$As$_2$(circles); (c):
K$_{0.7}$Nd$_{0.3}$Fe$_2$As$_2$(up-triangles); (d):
K$_{0.7}$La$_{0.3}$Fe$_2$As$_2$(down-triangles).}
\end{figure}

\begin{figure}
\includegraphics[width=0.50\textwidth]{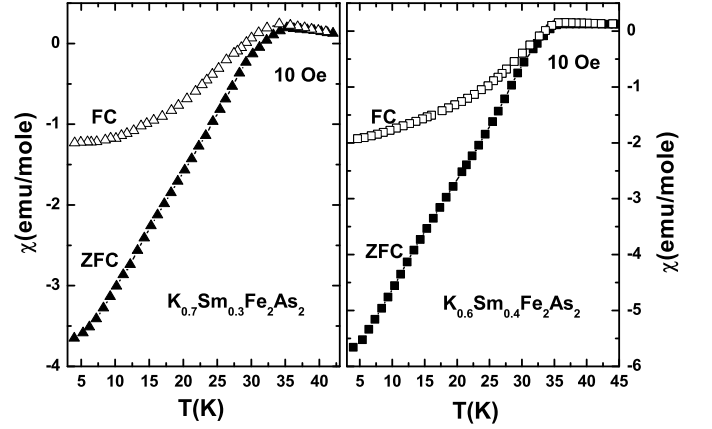}
\caption{Temperature dependence of susceptibility measured under 10
Oe in zero-field cooled and field cooled process for the samples,
(a): K$_{0.7}$Sm$_{0.3}$Fe$_2$As$_2$; (b):
K$_{0.6}$Sm$_{0.4}$Fe$_2$As$_2$.}
\end{figure}

 Figure 1 shows x-ray powder diffraction patterns for the samples
 K$_{0.7}$Ln$_{0.3}$Fe$_2$As$_2$ (Ln=Sm, Nd and La) and K$_{0.6}$Sm$_{0.4}$Fe$_2$As$_2$. All main peaks can
 be indexed by a tetragonal structure with a=0.3840 nm and c=1.3860 nm for the
 sample K$_{0.7}$Sm$_{0.3}$Fe$_2$As$_2$, a=0.3836 nm and c=1.3854 nm for the
 sample K$_{0.6}$Sm$_{0.4}$Fe$_2$As$_2$, a=0.3840 nm and c=1.3862 nm for the
 sample K$_{0.7}$Nd$_{0.3}$Fe$_2$As$_2$, a=0.3843 nm and c=1.3862 nm for the
 sample K$_{0.7}$La$_{0.3}$Fe$_2$As$_2$. The lattice constants a and c for all samples
 are slightly less than that of KFe$_2$As$_2$. A little of impurity
 LnAs and FeAs is observed in the x-ray diffraction pattern
 because K easily volatilizes and reacts with quartz
 tube at high temperature. In addition, annealing temperature of 740$\celsius$ may be
 too low for reacting sufficiently.

Figure 2 shows temperature dependence of resistivity for the samples
K$_{0.7}$Ln$_{0.3}$Fe$_2$As$_2$ (Ln=Sm, Nd and La) and
K$_{0.6}$Sm$_{0.4}$Fe$_2$As$_2$. The resistivity behavior for all
samples is similar to that observed in the samples
Ba$_{1-x}$K$_{x}$Fe$_2$As$_2$(x=0.3-0.7)\cite{hchen}. The
superconducting transition is observed at about 36 K and 35 K in the
samples K$_{1-x}$Sm$_{x}$Fe$_2$As$_2$(x=0.4) or x=0.3, respectively.
The samples K$_{0.7}$Ln$_{0.3}$Fe$_2$As$_2$ (Ln=Nd and La) show the
superconductivity at 34 K. We tried to increase the Ln-doping, but
large amount of impurity phase shows up. It suggests that more Ln
cannot induce into the system by conventional solid state reaction.

To confirm the superconductivity observed in resistivity for the
samples with nominal compositions K$_{1-x}$Sm$_{x}$Fe$_2$As$_2$
(x=0.3 and 0.4), the susceptibility measured under 10 Oe in
zero-field cooled and field-cooled cycle is shown in Fig.3. Figure 3
shows superconducting transition for both of the samples. The
transition temperature is about 35 K for the sample
K$_{0.7}$Sm$_{0.3}$Fe$_2$As$_2$ and 36 K for the sample
K$_{0.6}$Sm$_{0.4}$Fe$_2$As$_2$, respectively. These transition
temperatures are consistent with that observed in resistivity. The
data in Fig.3 give superconducting volume fraction of about 8\%
shielding fraction and about 3\% Meissner fraction at 5 K for sample
K$_{0.7}$Sm$_{0.3}$Fe$_2$As$_2$, and about 12\% shielding fraction
and about 4\% Meissner fraction for K$_{0.6}$Sm$_{0.4}$Fe$_2$As$_2$.
These results indicate the bulk superconductivity.

\begin{figure}[h]
\includegraphics[width=0.50\textwidth]{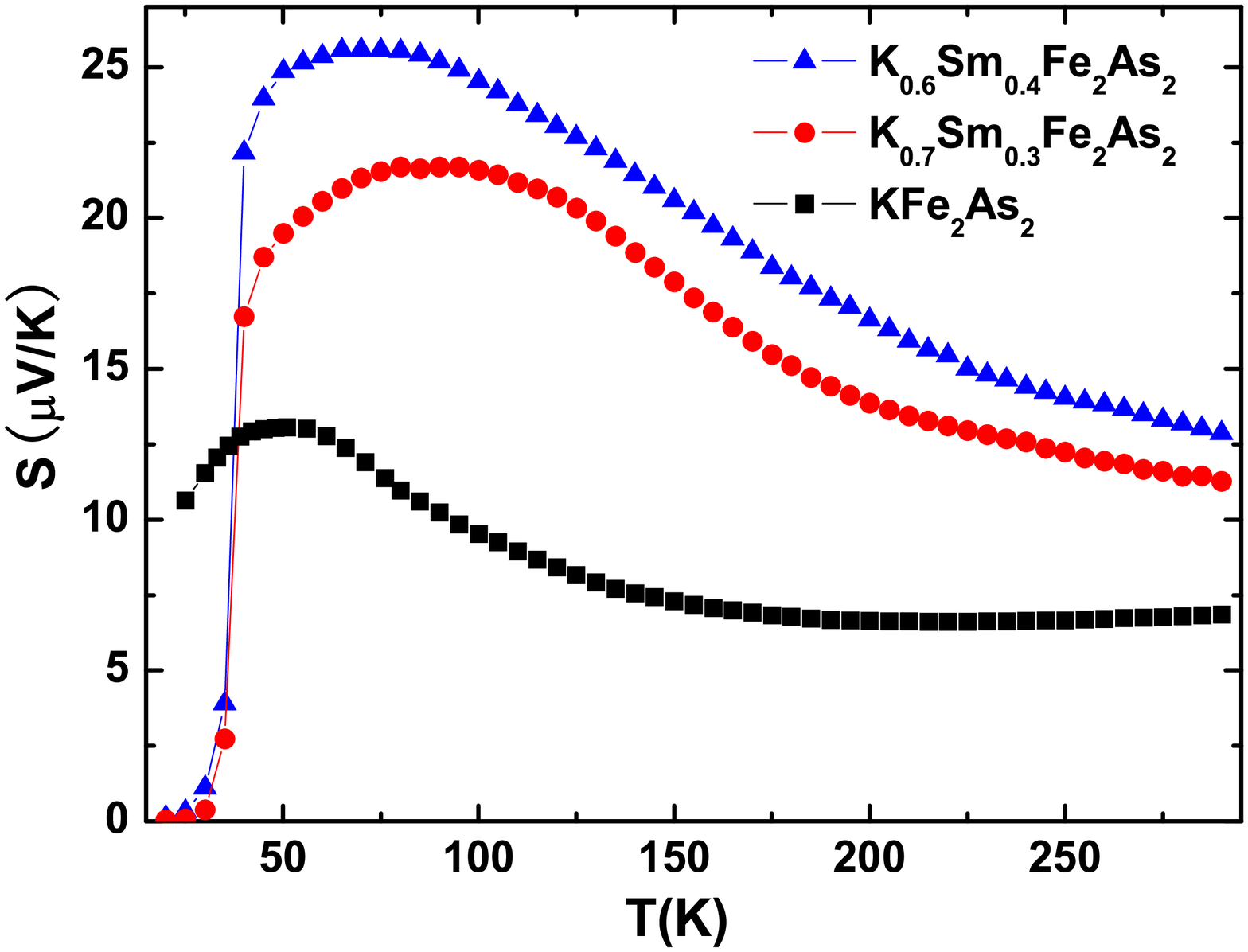}
\caption{Temperature dependence of thermoelectric power for the
samples (a): KFe$_2$As$_2$ (squares); (b):
K$_{0.7}$Sm$_{0.3}$Fe$_2$As$_2$ (circles); (c):
K$_{0.6}$Sm$_{0.4}$Fe$_2$As$_2$ (up-triangles).}
\end{figure}

In order to identify whether K$_{1-x}$Sm$_{x}$Fe$_2$As$_2$ (x=0.3
and 0.4) system is similar to the sample
Ba$_{0.6}$K$_{0.4}$Fe$_2$As$_2$ with p-type carrier, the
thermoelectric power (TEP) is systematically measured. Figure 4
shows the temperature dependence of thermoelectric power. TEP for
the sample KFe$_{2}$As$_{2}$ is positive, being different from that
for the parent compound BaFe$_{2}$As$_{2}$  and LaOFeAs
\cite{gwu1,mcguire}. Sm-doping enhances TEP magnitude,  but does not
change the sign of TEP. TEP for the samples
K$_{1-x}$Sm$_{x}$Fe$_2$As$_2$ (x=0.3 and  0.4) increases
monotonically with decreasing temperature, being similar to that
observed in the sample Ba$_{0.6}$K$_{0.4}$Fe$_2$As$_2$. From 75 K to
40 K, TEP decreases with decreasing temperature. At about 35 K, TEP
sharply decreases to zero, indicating a superconducting transition
which is consistent with those observed in resistivity and
susceptibility. One open question that one may ask whether
KFe$_2$As$_2$ is a real parent compound for the superconducting
K$_{1-x}$Ln$_{x}$Fe$_2$As$_2$ system. If one takes KFe$_2$As$_2$ as
a parent compound, the KFe$_2$As$_2$ is different from other parent
compounds: LnFeAsO (Ln=La-Gd) and $AEFe_2As_2$ (AE=Ba, Sr, Ca and
Eu) in which there exist structural and SDW instabilities, and
doping with electron and hole suppresses such instabilities, while
KFe$_2$As$_2$ shows superconductivity at about 3 K. In the
K$_{1-x}$Ln$_{x}$Fe$_2$As$_2$ (Ln=Sm, Nd and La) system, no
structural and SDW instabilities exist. It suggests that spin
fluctuation could be ignored. However, strong ferromagnetic and
antiferromagnetic fluctuations have been proposed to be responsible
for high-$T_c$ superconductivity in pnictides\cite{cao,dai,ma}.

In summary, the samples K$_{1-x}$Ln$_{x}$Fe$_2$As$_2$ (Ln=La, Nd and
Sm) with isostructural Ba$_{1-x}$K$_{x}$Fe$_2$As$_2$ were
systematically studied by resistivity, susceptibility, and
thermoelectric power(TEP). Substitution of Ln (Ln=La, Nd and Sm) for
K in $KFe_2As_2$ raises the superconducting transition temperature
to 34-36 K. The behavior of resistivity for the samples
K$_{1-x}$Ln$_{x}$Fe$_2$As$_2$ is similar to that observed in the
samples Ba$_{1-x}$K$_{x}$Fe$_2$As$_2$. The TEP measurements indicate
that the TEP of K$_{1-x}$Ln$_{x}$Fe$_2$As$_2$ is positive, being
similar to the samples Ba$_{1-x}$K$_{x}$Fe$_2$As$_2$ with p-type
carrier. In contrast to the Ba$_{1-x}$K$_{x}$Fe$_2$As$_2$ system, no
structural and SDW instabilities exist in the
K$_{1-x}$Ln$_{x}$Fe$_2$As$_2$ (Ln=Sm, Nd and La) system.

{\bf Acknowledgment:} This work is supported by the Natural Science
Foundation of China and by the Ministry of Science and Technology of
China (973 project No:2006CB601001) and by Natural Basic Research
Program of China (2006CB922005).

\end{document}